\documentclass[acmlarge]{acmart}

\AtBeginDocument{%
  }

\setcopyright{none}
\makeatletter
\renewcommand\@formatdoi[1]{\ignorespaces}
\makeatother
\renewcommand\footnotetextcopyrightpermission[1]{} % removes footnote with conference information in first column
\usepackage{caption}
\usepackage{subcaption}
\usepackage{graphicx}
\usepackage{hyperref}
\usepackage{multirow}
\usepackage{makecell}
\usepackage{hyperref}
\usepackage{soul}
\usepackage{mdframed}
\newmdtheoremenv{theo}{Theorem}
\usepackage[most]{tcolorbox}
\usepackage{booktabs}

\begin{document}

\title{\textit{Sens-BERT}: Enabling Transferability and Re-calibration of Calibration Models for Low-cost Sensors under Reference Measurements Scarcity}
\makeatletter
\let\@authorsaddresses\@empty
\makeatother
\author{M V Narayana}
\email{ee18d302@smail.iitm.ac.in}
\orcid{0000-0002-7081-3168}
\affiliation{%
  \institution{Department of Electrical Engineering, IITM}
  \streetaddress{ESB 219}
  \city{Chennai}
  \state{Tamil Nadu}
  \country{India}
  \postcode{600036}
}
\author{Kranthi Kumar Rachavarapu}
\affiliation{%
 \institution{Department of Electrical Engineering, IITM}
 \streetaddress{}
 \city{Chennai}
 \state{Tamil Nadu}
 \country{India}
 }
\author{Devendra Jalihal}
\affiliation{%
 \institution{EE Department, IITM}
 \streetaddress{}
 \city{Chennai}
 \state{Tamil Nadu}
 \country{India}
 }
\author{Shiva Nagendra S M}
\affiliation{%
 \institution{CE Department, IITM}
 \streetaddress{}
 \city{Chennai}
 \state{Tamil Nadu}
 \country{India}
 }

%
% By default, the full list of authors will be used in the page
% headers. Often, this list is too long, and will overlap
% other information printed in the page headers. This command allows
% the author to define a more concise list
% of authors' names for this purpose.
\renewcommand{\shortauthors}{Narayana et al.}

\begin{abstract}
Low-cost sensors (LCS) are becoming increasingly relevant for monitoring and understanding air quality at high spatial and temporal resolution. However, LCS measurements are noisy, which limits large-scale adaptability. Calibration is generally used to get good estimates of air quality measurements out from LCS. In order to do this, LCS sensors are typically co-located with reference stations for some duration. A calibration model is then developed to transfer the LCS sensor measurements to the reference station measurements. Existing works implement the calibration of LCS as an optimization problem in which a model is trained with the data obtained from real-time deployments; later, the trained model is employed to estimate the air quality measurements of that location. However, this approach is sensor-specific and location specific and needs frequent re-calibration. The re-calibration also needs massive data like initial calibration, which is a cumbersome process in practical scenarios.

To overcome these limitations, in this work, we propose \textit{Sens-BERT}, a BERT-inspired learning approach to calibrate LCS, and it achieves the calibration in two phases: self-supervised pre-training and supervised fine-tuning. In the pre-training phase, we train \textit{Sens-BERT} with only LCS data (without reference station observations) to learn the data distributional features and produce corresponding embeddings. We then use the \textit{Sens-BERT} embeddings to learn a calibration model in the fine-tuning phase. Our proposed approach has many advantages over the previous works. Since the \textit{Sens-BERT} learns the behavior of the LCS, it can be transferable to any sensor of the same sensing principle without explicitly training on that sensor. It requires only LCS measurements in pre-training to learn the characters of LCS, thus enabling calibration even with a tiny amount of paired data in fine-tuning. We have exhaustively tested our approach with the Community Air Sensor Network (CAIRSENSE) data set, an open repository for LCS. We show that the proposed method outperforms well-known calibration models such as single-variable linear regression, multiple-variable linear regression, and Random forest models.

\end{abstract}

\keywords{Low-cost sensors, Calibration, Representation Learning, BERT}


\maketitle
\thispagestyle{empty}
\fancyfoot{} 
\settopmatter{printacmref=False}

\section{Introduction}
Air pollution is a significant environmental concern that affects human health and well-being. Monitoring air quality is crucial for understanding the levels of pollutants in the atmosphere and taking appropriate actions to mitigate their effects. While professional-grade air quality monitoring (AQM) systems are widely used, they can be expensive and limited in availability \cite{Rai2017}. To address this issue, low-cost sensors (LCS) for AQM have emerged as a viable solution. LCS provide an affordable and accessible means of measuring various air pollutants, enabling individuals, communities, and organizations to monitor their local air quality actively \cite{Maag_wair, realtime_airquality, mobilemonitoing2019}. These sensors utilize innovative technologies and compact designs to deliver real-time data on pollutant concentrations, enabling timely responses and informed decision-making. LCS can measure parameters such as particulate matter ($PM$), gases like carbon dioxide ($CO_2$), carbon monoxide ($CO$), nitrogen dioxide ($NO_2$), and volatile organic compounds ($VOC$s). They may also include temperature ($T$) and humidity ($Rh$) sensors to assess environmental conditions comprehensively.

Despite these advantages, the measurements produced by the LCS are susceptible to various error sources that can affect the accuracy, rendering the LCS less reliable than the reference grade instruments. Typical error sources include variations in temperature and relative humidity, cross-sensitivities (sensitive towards pollutants that existed other than the target parameters) and drift \cite{Maag_survey, Narayana_establishing_2022}. Therefore, proper calibration is crucial to address these issues, and by doing so, we can improve the accuracy of LCS measurements \cite{feenstra_12lcs}. Calibration involves transforming the LCS measurements to align with reference station observations. 

Applying correction factors to the sensor signal for the respective parameters influencing its response is a preliminary way to calibrate the LCS \cite{Aurora_temp_correction, Nilson_temp_correction, Popoola_temp_corr, Mead_meter_ec}. For instance, $\mathcal{S}$ is an LCS signal, which corrects for the effect of temperature ($T$) and relative humidity ($Rh$) in the equation below.
\begin{equation}
\mathcal{\hat{S}} = \mathcal{S}+\alpha _1 (T) + \alpha _2 (Rh)
\end{equation}
where $\alpha _1 (.)$ and $\alpha _2 (.)$ are the correction factors modelled by exposing the LCS to various $T$ and $Rh$ values in controlled environments. The correction factors are sensor specific and often derived based on a limited number of calibration points. As a result, they may not accurately account for variations in the sensor's response across its entire operating range. At the same time, LCS may experience calibration drift over time due to environmental factors, sensor ageing, or changes in response characteristics which needs the frequent modelling of calibration factors, which is a cumbersome process in practice. 

To overcome the limitations in the correction factors approach, a learning-based approach is proposed to calibrate the LCS \cite{Spinelle_lr_mlr_ann, Zimmerman_rf, Cheng_ann}. In this approach, calibration of LCS is framed as an optimisation problem, where a calibration model $(f_\theta)$ is trained to minimise the mean square error (\textit{MSE}) between the LCS measurements ($\mathcal{X}_t$) and reference station observations ($\mathcal{Y}_t$). Since the calibration models are trained with ambient conditions in real-time deployments, they can handle the error sources effectively, provided they are trained on a substantial amount of the data that covers all these variables. However, in the existing learning-based calibration works, the calibration model is sensor-specific and location-specific since their optimization is limited to a particular scenario, and they do not generalize the characteristics of the LCS measurements. Due to this, their calibration accuracy is limited, and transferring the calibration models trained on one sensor to other sensors further limits their effectiveness in calibrating the Low-cost sensors. At the same time, it needs frequent re-calibration to handle the calibration drift, which requires substantial co-location measurements of LCS and reference stations.

In summary, the existing learning-based works mainly focus on transforming the LCS measurements without learning their distributional characteristics, such as temporal and environmental dependencies. It is essential to learn the distributional information of measurements to generalize the calibration model to the LCS and to improve the calibration accuracy with limited paired data. To address these limitations, we propose \textit{Sens-BERT},  a BERT-based (Bidirectional Encoder Representations from Transformers) \cite{bert} approach to calibrate the Low-cost sensors and it achieves the calibration in two steps: pre-training and fine-tuning. The pre-training step is self-supervised, where a BERT-based model, \textit{Sens-BERT}, is trained only on LCS measurements without using the reference station observations, and this pre-training enables \textit{Sens-BERT} to learn the characteristics of LCS measurement. Therefore, the pre-trained \textit{Sens-BERT} can produce the embeddings corresponding to the LCS measurements, which are then used in the fine-tuning step, where a calibration model can be trained with limited paired data. At the same time,  the pre-trained \textit{Sens-BERT} can be utilized in the re-calibration, and it can be transferred to other sensors of the same sensing principle that needs to be calibrated since it already learned the data characteristics.

To the best of our knowledge, this is the first work that implements a BERT-based approach to calibrate the LCS, and our contributions are as follows, 
\begin{itemize}
    \item We implement a BERT-based deep learning architecture, Sens-BERT, to learn the characteristics of LCS measurements, which enables calibration and re-calibration with limited paired data.   
    \item We empirically validate the transferability of \textit{Sens-BERT} to other sensors without explicitly pre-training it on those sensors. 
    % \textcolor{red}{Say a line about results,}
    \item We show that the proposed approach outperforms the existing optimization-based calibration works for different sensors in the CAIRSENSE data set \cite{cairsense_dataset}, an open repository for the LCS. 
    
\end{itemize}

\section{Related work} 
\label{sec:related_work}
Calibration approaches for low-cost sensors (LCS) have been an active area of research, and several works have been disseminated to enhance the accuracy of LCS through calibration. This section summarises the existing works on the calibration of LCS. 

\noindent\textbf{Laboratory and field studies.}
Numerous laboratory and field investigations have been undertaken to gather data from LCS deployed in diverse environments \cite{Li_evalu_nine,Ahn_labstudy,Kuula_labstudy,Lewis_evalution_chem}. This data collection enables examining sensor responses across different conditions and provides valuable inputs for developing calibration techniques. Concurrently, comparing LCS measurements with those obtained from established air quality monitoring stations facilitates the validation of LCS performance.

\noindent\textbf{Correction factors approach.} The data collected in both field and laboratory settings enables the development of correction factors for sensor response, effectively addressing biases in the sensor response. Carl et al. \cite{Malings_correc} and Barkjohn et al. \cite{Barkjohn_pmcorec} applied hygroscopic growth factor correction to PM sensor response in order to correct the effect of relative humidity on PM sensors. Similarly, the gas sensors also corrected for meteorological parameters to improve accuracy \cite{Popoola_temp_corr,Aurora_temp_correction, Aurora_correc}.

\noindent\textbf{Machine learning-based calibration.} Machine learning algorithms have been employed to calibrate LCS. The algorithm learns the relationship between the sensor responses and the actual pollutant concentrations by training the algorithm on data collected from LCS and corresponding reference instruments. This enables the algorithm to make more accurate predictions and calibrate the sensor readings. These models include single-variable linear regression\cite{Spinelle_calibration, Hasenfratz_lr, Saukh_lr, lin_lr}, multiple-variable linear regression\cite{KUMAR_sl_ml_knn,Spinelle_calibration, VanZoest_mlr}, K-nearest neighbours\cite{KUMAR_sl_ml_knn}, random forest\cite{Zimmerman_rf, wang_rf, Han_sl_ml_rf_lstm, narayana_quanti}, artificial neural networks\cite{Badura_sl_ml_ann, Han_sl_ml_rf_lstm, Cheng_ann, Alhasa_ann, Spinelle_lr_mlr_ann}, support vector machine \cite{Bigi_ml_svr_rf, Mahajan_svm}, Extreme Gradient Boosting\cite{Si_xgboost}, generalised additive model \cite{Munir_lr_gam}. In addition, a few works explored hybrid models obtained by combining two or more ML to calibrate the LCS \cite{Lin2018, Hagan_hybrid,Malings_hybrid}. For instance, Hagan et al. \cite{Hagan_hybrid} and Lin et al. \cite{Lin2018} combined the regression model with the k-nearest neighbours and random forest, respectively, to improve the calibration accuracy compared to the stand-alone models. A similar approach that combines regression models with ANN was implemented by  Cordero et al. \cite{Cordero_hybrid}; instead of connecting the models, Ferrer-Cid et al. used multi-sensor data fusion techniques to calibrate the LCS \cite{Ferrer-Cid_data_fusion}.

In the existing machine-learning-based calibration works, the calibration of low-cost sensors is not transferable to other sensors since they learn from sensor-specific data alone instead of the general characteristics of LCS. Therefore, it requires a substantial amount of real-time deployment data for every recalibration, which is necessary to overcome the calibration drift problem. In contrast, \textit{Sens-BERT}, proposed in this work, first learns the data representation from low-cost sensors measurements alone, which enables the transfer of the \textit{Sens-BERT} to other sensors. At the same time, it can be calibrated with limited data for real-time deployments.   

\section{Approach}
This work aims to learn the representations of low-cost sensors (LCS) measurements with BERT architecture and needs to calibrate the LCS with limited paired data in real-time deployments. To accomplish the task, we first frame the calibration of LCS as a BERT-based deep learning model (\textit{Sens-BERT}) optimization (\textcolor{blue}{Sec \ref{subsec:prob_def}}). Then we cover the characteristics of LCS measurements (\textcolor{blue}{Sec \ref{subsec:data_chara}}), which helps to tailor the input to the \textit{Sens-BERT}. Finally, we describe the architecture of Sens-BERT in \textcolor{blue}{Sec \ref{subsec:sens_bert}}.
% fig 1 calibration of LCS as an optimization problem
\begin{figure}[hbt!]
    \centering
    \includegraphics[width=0.7\linewidth]{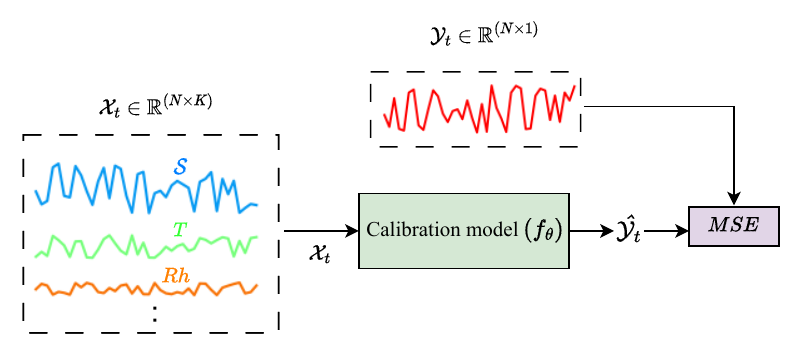}
    \caption{Calibration of LCS as an optimization problem in the existing works}
    \label{fig:existing_datadriven}
\end{figure}

\subsection{Problem formulation: implementation of calibration of LCS as a BERT-based deep learning model} \label{subsec:prob_def}

In low-cost sensor calibration, the goal is to transform the LCS measurements to reference station observations. Specifically, given a set of measurements of LCS and reference station observations, $\{\mathcal{X}_t, \mathcal{Y}_t\}^T_{t = 1}$,  our objective is to learn a regression model that transforms $\mathcal{X}_t\in\mathbb{R}^{N\times K}$ to $\mathcal{Y}^{N\times 1}_t$. Here, $\mathcal{X}_t$ consist $N$ of measurements from each $K$ low-cost sensors such as temperature ($T$), relative humidity ($Rh$), pollutant of interest like $PM_{2.5}$, $CO_2$ etc.,  and $\mathcal{Y}_t$ consists corresponding reference station observations. Traditionally, this problem is approached as a learning-based optimization problem, as shown in Fig. \ref{fig:existing_datadriven}, where the parameters ($\theta$) of a calibration model ($f_\theta$) are optimized such that it can transform $\mathcal{X}_t$ to $\mathcal{Y}_t$, as shown in equations (\ref{eq:calibration_model}) and (\ref{eq:theta_optimization}).
\begin{align} 
    \mathcal{\hat{Y}}_t &= f_{\theta}(\mathcal{X}_t) \qquad\forall t \label{eq:calibration_model} \\
    \theta^* &= \arg \min_{\theta} \frac{1}{N}\sum_{t=1}^{N}MSE(\mathcal{\hat{Y}}_t,\mathcal{Y}_t) \label{eq:theta_optimization}
\end{align}

However, the measurements in $\mathcal{X}_t$ show temporal dependency, which means the current observations are greatly influenced by previous measurements \cite{temporaldependency_zheng}, and these temporal dependencies cannot be tackled by the traditional approaches. Therefore, it needs to be hypothesised that the calibration model ($f_\theta$) can learn the temporal dependencies and other distributional information of $\mathcal{X}_t$ to improve calibration accuracy, as shown in equation (\ref{eq:temporal_depen}). 
\begin{align} \label{eq:temporal_depen}
     \mathcal{\hat{Y}}_t &= f_{\theta}(\mathcal{X}_{t-M},\mathcal{X}_{t-M-1}, .. \, \mathcal{X}_t) \qquad\forall t  
\end{align}
Where $M$ is the temporal context slice which is the experimental hyperparameter. Since the calibration model $f_\theta$ in equation (\ref{eq:temporal_depen}) focuses only on transforming $\mathcal{X}_t$ to $\mathcal{Y}_t$ without learning the temporal dependencies and distributional information, it needs to alter the $f_\theta$ such that it can learn the characteristics of the measurements first and then it can transform the LCS measurements to reference station observations. 

\begin{figure}[hbt!]
    \centering
    \includegraphics[width=\linewidth]{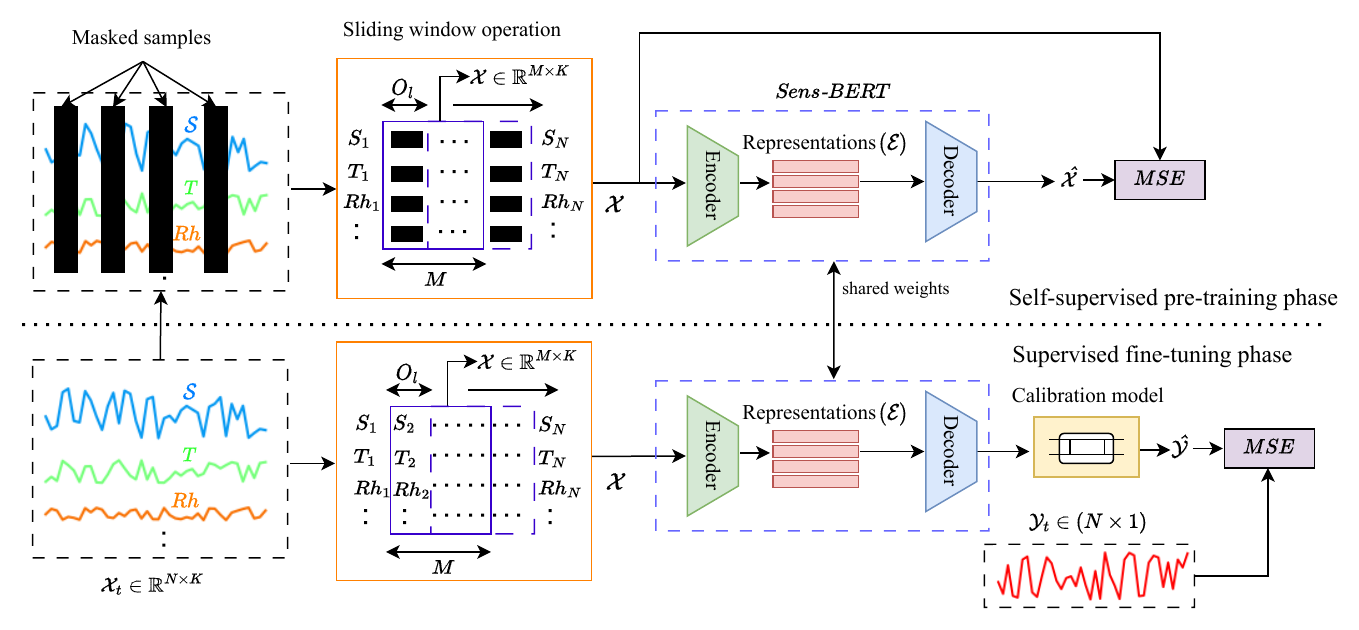}
    \caption{Calibration of Low-cost sensors with \textit{Sens-BERT}}
    \label{fig:sens_bert}
\end{figure}

 We propose a two-step calibration framework shown in Fig. \ref{fig:sens_bert} to implement such a calibration process. In the proposed approach, we first focus on learning the temporal dependencies and other distributional information of $\mathcal{X}_t$ in the self-supervised pre-training step using a BERT-based deep learning architecture, \textit{Sens-BERT}. The \textit{Sens-BERT} shown in Fig. \ref{fig:sens_bert} inherits an encoder and decoder to learn the characteristics of LCS measurements. To force the \textit{Sens-BERT} to learn the characteristics of LCS measurements, we mask sequences of samples in $\mathcal{X}_t$ using the span masking technique \cite{Xu_limubert} with probability $P$ and then operated a sliding window of length $M$ with an overlapping length of $O_l$ between two successive windows such that it can feed the \textit{Sens-BERT} with temporal context. Let $\mathcal{X} \in \mathbb{R}^{M \times K}$ is a sample obtained by operating sliding window on $\mathcal{X}_t$ at time $t$, then the encoder transforms the measurements in $\mathcal{X}$, including masked samples, into embeddings ($\mathcal{E}$), and the decoder re-transforms $\mathcal{E}$ to $\mathcal{\hat{X}}$, which is the estimated $\mathcal{X}$. Therefore, in the pre-training phase, \textit{Sens-BERT} transforms $\mathcal{X}$ to $\mathcal{E}$ and then to $\mathcal{\hat{X}}$ as shown in equations (\ref{eq:encoder}) and (\ref{eq:decoder}) and learns the characteristics of $\mathcal{X}_t$ while doing these transformations, by minimizing the loss in equation (\ref{eq:loss_pretrain}), that is, the $MSE$ between $\mathcal{X}$ and $\mathcal{\hat{X}}$. 

\begin{align}  
\mathcal{E}& = f_{enc}(\mathcal{X}) ; \quad  \mathcal{X}  \in \mathbb{R}^{M\times K} \sim \{\mathcal{X}_t\}^T_{t=1} \quad \forall t  \label{eq:encoder}\\
 \mathcal{\hat{X}} &  = f_{dec}(\mathcal{E}) \label{eq:decoder}\\
 \text{loss}_{\text{pre-train}} & =  \frac{1}{N}\sum_{i=1}^{N}MSE(\mathcal{X}, \mathcal{\hat{X}}) \label{eq:loss_pretrain}
\end{align}

Once Sens-BERT is trained in pre-training, it can be used in the second step, which is a fine-tuning phase, where a calibration model ($f_\theta$) is trained with generated embedding against the reference station observations $\mathcal{Y}_t$, as shown in the below equations.
\begin{align} \label{eq:finetune}
    \mathcal{\hat{X}} & = \text{\textit{Sens-BERT}}(\mathcal{X}) \qquad \forall t  \\
    \mathcal{\hat{Y}}_t & = f_{\theta}(\mathcal{\hat{X}} )\\
    \text{loss}_{\text{fine-tune}} & = \frac{1}{N}\sum_{t=1}^{N}MSE(\mathcal{\hat{Y}}_t-\mathcal{Y}_t)
\end{align}

Since our approach learns the temporal dependencies and other distributional information of the LCS measurements, it can calibrate the LCS effectively with limited paired data. At the same time, the pre-trained \textit{Sens-BERT} can be transferred to other sensors, which need to be calibrated.

\subsection{Characteristics of LCS data } \label{subsec:data_chara}

% As mentioned, Sens-BERT is implemented on BERT architecture, a language processing model designed to process text, which differs from numerical sensor data processing. In the case of language processing, input to the BERT is one-hot vectors of the context (words or sentences), but in LCS calibration, it is vectors of absolute values. Therefore, knowing the variables that need to be considered in learning LCS characteristics is crucial. In order to customize input data set that is used to train the \textit{Sens-BERT}, we have studied the characteristics of LCS data from the literature and obtained the following observations.

\begin{figure}[hbt!]
    \begin{minipage}{0.5\linewidth}
        \includegraphics[width = \linewidth, trim=0 0.1cm 0 0,clip]{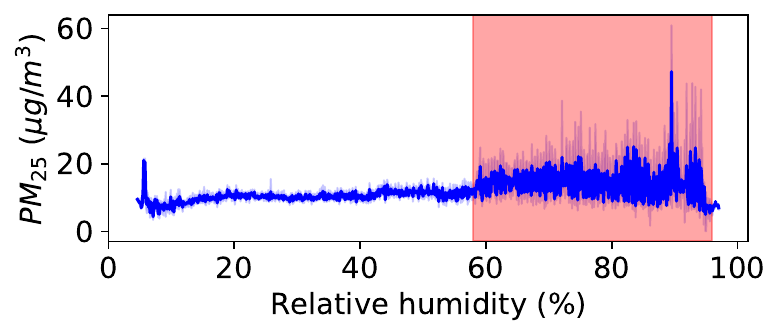}\\
        \label{fig:meterology_a}
        \subcaption{ Relative humidity effect on particle sensors}
    \end{minipage}%
    \begin{minipage}{0.5\linewidth}
        \includegraphics[width = \linewidth, trim=0 0.1cm 0 0,clip]{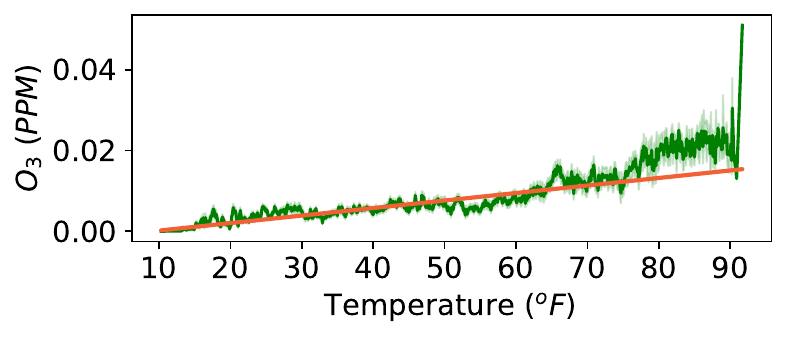}\\
        \subcaption{Temperature effect on metal oxide gas sensors)}
    \end{minipage}%
\caption{Effect of temperature and relative humidity on the response of low-cost sensors. The plots are drawn with the data from the CAIRSENSE data set \cite{cairsense_dataset}.}
\label{fig:effect_meterology}
\end{figure}

Generally, low-cost sensors predominant in AQM are divided into particle sensors that work on the light scattering principle and gas sensors that work on metal oxide or electro-chemical principle. Particle sensors are greatly influenced by relative humidity ($Rh$) due to the hygroscopic properties of the particulate matter ($PM$) \cite{Malings_hybrid}. At higher relative humidity, tiny particles floating around (aerosols) tend to stick together more because of the water vapour, which significantly changes the amount of scattered light, thus changing the response of the sensors. To illustrate the effect of relative humidity on particle sensors, we plotted sample data of AirAssure, an LCS that works on the light scattering principle from the CAIRSESE data set \cite{cairsense_dataset} in Fig. \ref{fig:effect_meterology}(a). The graph is plotted between the $\text{PM}_{25}$ ($PM$ of size less than 0.25 $\mu m$) on the y-axis and $Rh$ on the x-axis. From Fig. \ref{fig:effect_meterology}(a), it can be observed that the $\text{PM}_{25}$ values have evident fluctuations for higher $Rh$, which is highlighted in red shade.

Temperature ($T$) and relative humidity ($Rh$) are the prime environmental factors influencing the accuracy of gas sensors. These factors influence the resistance of the heating element under the metal oxide surface in metal-oxide sensors; thus, changing its sensitivity causes inaccurate measurements \cite{Aurora_temp_correction, Masson_rh_effect_mox}. Fig. \ref{fig:effect_meterology}(b),  plotted between the temperature and response of Aeroqual, a metal oxide sensor measures $O_3$, shows the apparent drift (drift in trend line) in the response with increase in temperature. In the case of electrochemical sensors, temperature changes the rate of the chemical reaction and moisture content depletes the sensors' electrodes, thus changing the response of the sensors \cite{Mead_meter_ec}. Further, they suffer from cross-sensitive issues, which means they are also sensitive to other gasses in the atmosphere  \cite{Liu_gassensors, maag_predep}.

Based on the above observations, we select the variables that need to be considered in the calibration implementation, discussed in Sec. \ref{sec:experiments}.

\subsection{Sens-BERT} \label{subsec:sens_bert}
\textit{Sens-BERT}, shown in Fig. \ref{fig:sens_bert}, is implemented on BERT architecture, consisting of an encoder and a decoder. Encoders transform the given sequences ($\mathcal{X}$) into embeddings ($\mathcal{E} \in \mathbb{R}^{H_{dim} \times M}$), and the decoder re-transforms the embeddings into sequences, as represented in the below equations.
\begin{align*} 
\boxed{
\begin{array}{rcl}
& \textbf{\textit{Sens-BERT}}\\
& \qquad \mathcal{X} \quad  \sim \quad \{\mathcal{X}_t\}^T_{t=1} \quad \forall t \\
& \mathcal{E} \quad = \quad f_{enc}(\mathcal{X})\\
 & \mathcal{\hat{X}} \quad = \quad f_{dec}(\mathcal{E})
\end{array}
}
\end{align*}

% \begin{equation*}
% \boxed{
% \begin{array}{rcl}
% & & \textbf{\textit{Sens-BERT}}\\
% \blue{(\mathcal{X},\mathcal{Y}_t) & \sim &\{(\mathcal{X}_t,\mathcal{Y}_t)\}^T_{t=1} \quad \forall t } \\
% \mathcal{E}& = &f_{enc}(\mathcal{X}) ; \quad \mathcal{X}  \in \mathbb{R}^{M\times K} \sim \{\mathcal{X}_t\}^T_{t=1} \quad \forall t  \\
% \mathcal{\hat{X}}  & = & f_{dec}(\mathcal{E})
% \end{array}
% }
% \end{equation*}

 While doing these transformations, it learns the characteristics of low-cost sensor measurements by minimizing the loss equation in (\ref{eq:loss_pretrain}).
%  \begin{align} \label{eq:bert_loss}
%     \text{loss}_{\text{pre-train}} & =  \frac{1}{N}\sum_{i=1}^{N}MSE(\mathcal{X}, \mathcal{\hat{X}}).
% \end{align}
 The \textit{Sens-BERT} that learned the characteristics of LCS can be utilised to calibrate the LCS with limited paired data. At the same time, it can be transferable to other sensors that need a fresh calibration without explicitly training the \textit{Sens-BERT} on those sensors. Calibration of LCS with \textit{Sens-BERT} can be achieved in two steps -
\begin{enumerate}
    \item \textit{Self-supervised pre-training} to learn good representations using low-cost sensor measurements alone.
    \item \textit{Supervised fine-tuning}  to learn the final calibration model.
\end{enumerate}
\begin{figure}[hbt!]
    \centering
    \includegraphics[scale=0.7]{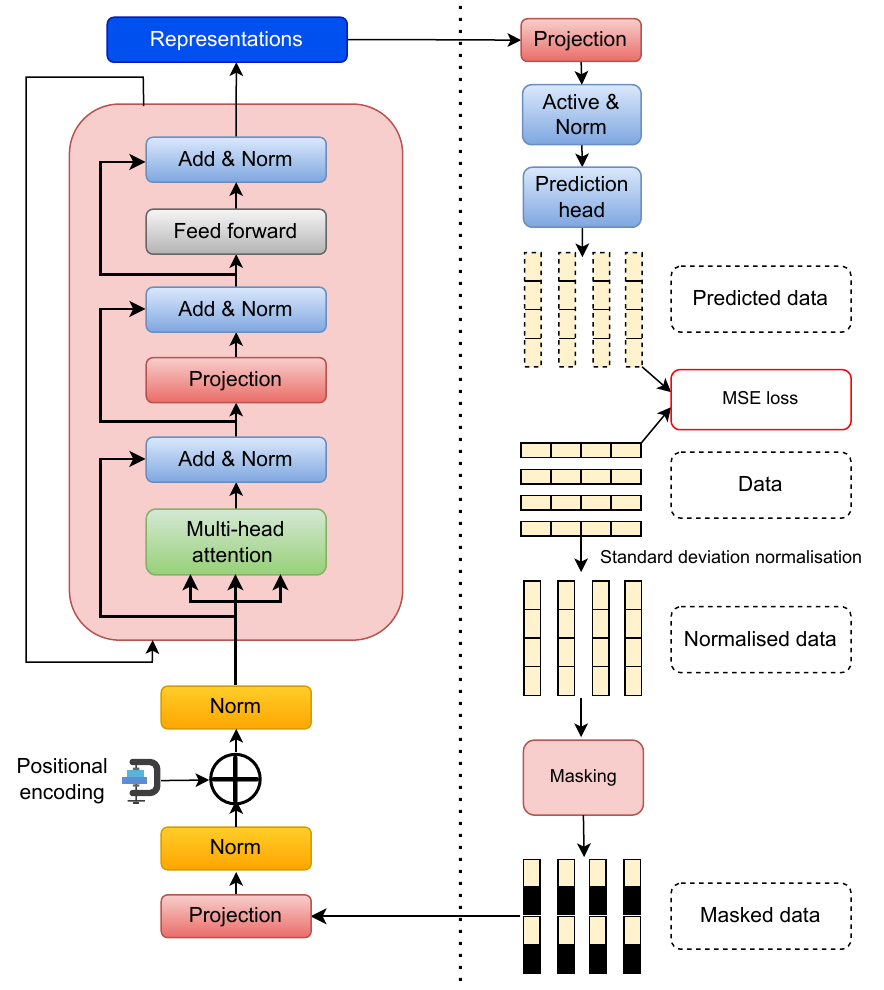}
    \caption{Self-supervised pre-training phase of Sens-BERT with transformer architecture}
    \label{fig:self_supervised_learning}
\end{figure}
\subsubsection{Self-supervised pre-training phase} \label{subsubsec:pre-train}
The objective of the pre-training phase is to learn the representations of  LCS measurements ($\mathcal{X}_t$) with \textit{Sens-BERT}, and the sequence of steps involved in the pre-training phase are illustrated in Fig \ref{fig:self_supervised_learning}. To force \textit{Sens-BERT} to learn the representations of $\mathcal{X}_t$, we first mask the sequence of samples in $\mathcal{X}_t$ with probability $P$ by using the span masking technique \cite{Xu_limubert}. Next, we operated a sliding window of size $M$ on $\mathcal{X}_t$ as shown in equation (\ref{eq:sliding_window}), to slice M samples from $\mathcal{X}_t$ at every time t. Where $M$ is the temporal context slice, an experimental parameter, $K$ is the number of variables (parameters considered for calibration) in $\mathcal{X}_t$ and $N$ is the number of samples. The sliding window operation enables to pass temporal context to the \textit{Sens-BERT}, so it can learn the temporal dependencies in the measurements. 
\begin{align} \label{eq:sliding_window}
    \mathcal{X} \in \mathbb{R}^{M \times K} \sim \{\mathcal{X}_t\}^T_{t=1} \quad; \mathcal{X}_t \in \mathbb{R}^{N\times K} , \mathcal{X} = \{\mathcal{X}_{t-M}, \mathcal{X}_{t-M-1} .. \, \mathcal{X}_t\} \forall t
\end{align}

Then we apply projection (\texttt{\textbf{Proj}(.)}) operation to the input sequences in  $\mathcal{X}$. Since the number of features or variables ($K$) in $\mathcal{X}$ is tiny to feed to the BERT architecture \cite{transformers}, that is, the encoder and decoders of \textit{Sens-BERT}, it needs to project the measurements in $\mathcal{X}$
to higher dimensional space, and we implemented this by using a linear layer shown in equation (\ref{eq:proj}).
\begin{align} 
\label{eq:proj}
    D = \texttt{\textbf{Proj}}(\mathcal{X}) = \mathbf{W}^{T} \mathcal{X}
\end{align}
Where $\mathbf{W}$ is a weight matrix of size $H_{dim}\times K$ and $H_{dim}$ is the hidden dimension size, the dimension expansion parameter of the neural networks. Once the values in $\mathcal{X}$ transform to higher dimensional space $D$, the vectors in D need to be normalized to train the transformer-based architectures effectively \cite{layer_noram}. We applied the layer normalization technique proposed by Jimmy et al. \cite{layer_noram} shown in equation (\ref{eq:layer_nor}) in our experimentation to normalize the values in $D$.  
\begin{align}\label{eq:layer_nor}
    D^{'}_{ji} = \texttt{LayerNorm}(D) = \frac{D_{ji}-\mu_{j}}{\sqrt{(\sigma^2_j+\epsilon)}}.\gamma+\beta
\end{align}
Where $\mu_j$ and $\sigma_j$ are the mean and standard deviation of $j^{th}$ column of D, $i$ is the row number and $\epsilon$ is a small number to eliminate the occurrence of infinite for zero value of $\sigma$. $\gamma$ and $\beta$ are the learning hyperparameters that help neural networks learn the dynamical normalization of features. Note that where ever the projection and normalization come in pre-training, it does the same in equations (\ref{eq:proj}) and (\ref{eq:layer_nor}).

Once the values in $D$ are normalized, it is added with the positional encoding with the help of a positional embedding function $PE(.)$  \cite{transformers}, which maps the column ($j$) index of D to a vector of length $H_{dim}$. Positional encoding helps to preserve the order of information that can be utilized at the decoder. Then, the data is passed through second-layer normalization, and the output after second-layer normalization, that is, the hidden features, are expressed as follows:
\begin{align}
    \mathcal{H}._j & = \texttt{LayerNorm}(D^\prime._j + \texttt{PE}(j)) 
\end{align}

Then, the attention-enteric block shown in Fig. \ref{fig:self_supervised_learning} (big pink rectangular box)  takes $H$ as input and repeats for $R_{num}$ of times before producing the final representations $\mathcal{E}$. It performs the two primary operations, Multi-head attention (\texttt{\textbf{Mult-Attn}(.)}) and feed-forward (\texttt{\textbf{FeedFrd}(.)}) operations, with projection and layer normalization after every operation. The \texttt{\textbf{Mult-Attn}(.)} is the self-attention layer of the transformer architecture \cite{transformers} with multiple attention heads. Here, we employ the scaled dot-product attention mechanism to learn the latent interaction within the embeddings, which can be considered as transforming three sets of LCS measurements, i.e., Query, Key, and Value representation. Here, the attention score is generated from Query and Key. Then, the Value vector is weighted according to the attention score to generate the final embedding.
% It transforms the words, in our case, the LCS measurements, in the input sequence into Query, Key, and Value representations and assigns an attention score to them, \textcolor{blue}{which are the embeddings.  
The \texttt{\textbf{FeedFrd}(.)} operation enables the element-wise non-linearity transformation of incoming embeddings.% to the transformer in the feedback from the second iteration.} 
We implemented the feed-forward operation using two fully connected layers coupled with a  Gaussian Error Linear Unit (GELU) \cite{gelu_activation} activation function. 
\begin{equation}
    \label{eq:attention_enteric}
 \{ \mathcal{H} = \texttt{LayerNorm(\textbf{FeedForward}(\texttt{LayerNorm(\textbf{Proj}(\texttt{LayerNorm(\textbf{Multi-Attn}($\mathcal{H}$))))}}} \} ^{R_{num}} = \mathcal{E} \in \mathbb{R}^{H_{dim} \times M } 
\end{equation}
Finally, the embeddings produced by the attention-entric block are passed to the decoder to reconstruct the masked samples. The decoder implementation involves a projection layer, an activated and normalization layer and a prediction head. At first, the projection layer projects the embeddings, $\mathcal{E}$, into D. Then the prediction layer (\texttt{\textbf{Pred(.)}}), followed by \texttt{LayerNorm(.)} operation, reconstructs $\mathcal{\hat{X}}$ from the D.
\begin{align} \label{eq:proj_decoder}
     D & = \texttt{Proj}(\texttt{\textbf{GELU}}(\mathcal{E})) \\
     \mathcal{\hat{X}} & = \texttt{LayerNorm}(\texttt{\textbf{Pred}}(D))
\end{align}

Putting all the operations together, the \textit{self-supervised pre-training phase} transforms $\mathcal{X}$ into $\mathcal{E}$ then to $\mathcal{\hat{X}}$ and makes \textit{Sens-BERT} learn the characteristics of low-cost sensor measurements while doing these transformations by minimizing the loss in the below equation.

$$\text{loss}_{\text{pre-train}} =  \frac{1}{N}\sum_{i=1}^{N}MSE(\mathcal{X}, \mathcal{\hat{X}})$$

\subsubsection{Supervised fine-tuning phase}
\begin{figure}[!hbt]
    \centering
    \includegraphics[scale=0.8]{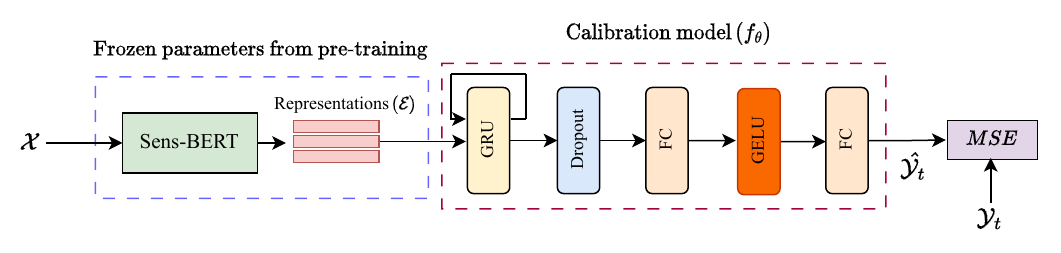}
    \caption{Supervised fine-tuning phase of calibration model with trained \textit{Sens-BERT}}
    \label{fig:supervised_lear}
\end{figure}
The trained Sens-BERT with frozen parameters can be utilized in the supervised fine-tuning phase, where a calibration-specific model can be trained with the limited paired data. The calibration-specific model is of user interest, and we implemented this by stacking three Gated Recurrent Unit (GRU) layers \cite{GRU}, one drop-out layer and two fully connected layers one on another, as shown in Fig. \ref{fig:supervised_lear}. 
% \textcolor{red}{needs to write the use of GRU, dropout and fully connected layers and how they convert $\mathcal{E}$ to $\mathcal{\hat{Y}}$}
Specifically, for given observations $\{\mathcal{X}, \mathcal{Y}_t\}$, where $\mathcal{X} \in \mathbf{R}^{M\times K}  \sim \{\mathcal{X}_t\}^T_{t=1}$ at time $t$,  is the low-cost sensor measurements from $K$ sensors over $M$, $\mathcal{Y}_t\in\mathbf{R}$ is the reference station measurement, we first extract the embeddings using the pre-trained sens-bert encoder as,
\begin{align}
    \mathcal{E}_t& = f_{enc}(\mathcal{X}_t) \in \mathbf{R}^{M\times d},
\end{align}
where $d$ is the embedding dimension. This embedding is then fed as input to the calibration model ($f_\theta$) to predict the corrected sensor measurement $\mathcal{Y}_t$ as,
\begin{align}
    \mathcal{\hat{Y}}_t & = f_{\theta}(\mathcal{E}_t ) = \texttt{FC}(\texttt{Dropout}(\texttt{GRU}(\mathcal{E}_t))).
\end{align}

Finally, the calibration-specific model is trained by minimizing the loss in equation (\ref{eq:loss_finetune}), which can be used to obtain accurate pollution values from LCS.

\begin{align} \label{eq:loss_finetune}
    \text{loss}_{\text{fine-tune}} & =  \frac{1}{N}\sum_{i=1}^{N}MSE(\mathcal{Y}_t, \mathcal{\hat{Y}}_t).
\end{align}

Therefore, $\textit{Sens-BERT}$ learns the characteristics of LCS measurements in the pre-training phase, which can be used to train a calibration model with limited paired data in the fine-tuning phase. Further, the trained \textit{Sens-BERT}  can be transferred to other sensors needing fresh calibration. The implementation details on the real-time data set are presented in Sec. \ref{subsec:implementation}, and the corresponding results are available in Sec. \ref{subsec:results}

% \subsubsection{Inference}

\section{Experiments} \label{sec:experiments}
\subsection{Data sets} \label{subsec: datasets}
We evaluate our approach on Community Air Sensor Network (CAIRSENSE) data set, an open repository developed by Feinberg et al. \cite{cairsense_dataset} for low-cost sensors, and it is available at the US environmental protection (USEPA) data website \cite{us_epa}. The data set contains measurements from various low-cost sensors and the time corresponding temperature, relative humidity and reference satiation observations. The measurements are taken from the Colorado state at a one-minute resolution.

Details of the low-cost sensors considered from the CAIRSENSE data set are as follows,
\begin{itemize}
    \item Shinyei - measures PM - works on light scattering principle
    \item SpecK -  measures PM - work on light scattering principle
    \item OPCPMF -  measures PM - works on light scattering principle
    \item TZOA -  measures PM - work on light scattering principle
    \item AirAssure -  measures PM - work on light scattering principle
\end{itemize} 
% We first, pre-train the \textit{Sens-BERT} with the measurements of the above-mentioned sensors separately, which facilitates a trained \textit{Sens-BERT} for each sensor. Then we use the trained \textit{Sens-BERT} to fine-tune the transformer-based calibration model with limited pair data. Then the fine-tuned calibration models are tested with test data. 
\subsection{Implementation details} \label{subsec:implementation}
As discussed in Sec. \ref{subsec:prob_def}, calibration models aim to transform the raw sensor signal ($\mathcal{S}_t$) to the corresponding reference station observations ($\mathcal{Y}_t$). Though it seems like handling a single variable ($\mathcal{S}_t$) transformation, in practice, it needs to be handled more than one variable since the LCS performance is severely affected by other covariates discussed in Sec \ref{subsec:data_chara}. Therefore, we framed the data set ($\mathcal{X}_t$) by including all the variables ($v_K; K =\{1, 2, ..\})$ that affect the performance of the selected sensor as;
\begin{align*}
    \mathcal{X}_t & = \begin{bmatrix}
        v_1 & v_2 & ..\;, v_K \\
    \end{bmatrix}_{N\times K} 
\end{align*}
Where $K$ is the number of variables, N is the number of samples and $v_k \in \mathbf{R}^{N \times 1}$ is the columns vector of measurements from $K^{th}$ variable.  In our experiments in Sec. \ref{subsec:results}, we consider output from the LCS ($\mathcal{S}$), temperature ($T$) and relative humidity $Rh$ as the variables in $\mathcal{X}_t$. $\mathcal{Y}_t \in \mathbf{R}^{N \times 1}$ are the time corresponding reference station observations to that of $\mathcal{X}_t$. Point to be noted that the $K$ value depends on the sensor that needs to be calibrated and influencing factors in the sensor deployment site.

% Since there exist studies on the effect of meteorology on the performance of LCS, including sensors that work on the light scattering principle \cite{Malings_correc, Zheng_correction}, as a cautionary note, we tested all the models with different combinations of variables ($v_1, ..\, v_K$) subsets and the results are presented in Sec. \ref{subsec:results}. 

\textit{Elimination of outliers}: We apply threshold-based filtering to discard the outlier samples. Since the air pollutants follow skewed distributions \cite{narayana_quanti}, we used the Inter-Quartile Range ($ IQR = Q3-Q1 $) proximity rule to detect the outliers; that is, the data points that fall below $Q1-1.5 \times IQR$ or above $Q3+ 1.5\times IQR$ are outliers, where $Q1$ and $Q3$ are the $25^{th}$ and $75^{th}$ percentile of the data, respectively.

\textit{Creation of overlapping sequences}: The measurements in $\mathcal{X}_t$ show temporal dependency, which means the current observations are greatly influenced by previous measurements \cite{temporaldependency_zheng}. To feed the temporal context to the \textit{Sens-BERT}, we operated a sliding wind of length ($M$) 128 with overlapping ($O_l$) of $M-1$ samples between two successive windows on $\mathcal{X}_t$. That means at time $t$; we slice 128 samples in $\mathcal{X}_t$ from $t$ to $t-M$ to train \textit{Sens-BERT} and we iterate this sliding window till the end of samples in $\mathcal{X}_t$. Here, $M$ and $O_l$ are the experimental hyperparameters. 

% \textit{Data normalization:}
% The BERT models generally do not consider data normalization since the inputs are well-normalized as one-hot vectors. But in our case, the inputs are the absolute values from different sensors, which shows significant differences in their distributions. In order to avoid the unwanted heterogeneity in data, we have normalized all the variables in the data set with the mean and standard deviation normalization (\textbf{SDN}) as shown in the below equation where $\mu$ and $\sigma$ are the mean and standard deviation of variable $v$. 
% \begin{equation} \label{eq:normaliztion}
%     \text{\textbf{SDN} ($v_j$)} = \frac{v_j-\mu}{\sigma}
% \end{equation}

 \textit{Masking of samples}: To ensure \textit{Sens-BERT} learns  characteristics of $\mathcal{X}_t$ in pre-training, we have masked sequence of samples in $\mathcal{X}_t$ using span masking technique \cite{Xu_limubert} with probability ($P$) of 0.2. The length of subsequent samples ($S_l$) to mask and $P$ are the experimental parameters.

% If $\mathcal{X}_t$ and $\mathcal{Y}_t$

\textit{Training and testing}:
We divide the data in $\{\mathcal{X}_t, \mathcal{Y}_t \}$ into training data (\{$\mathcal{X}_t^{train}, \mathcal{Y}_t^{train}\}$) and testing data  (\{$\mathcal{X}_t^{test}, \mathcal{Y}_t^{test}\}$) in 0.9 and 0.1 ratios, respectively. We then pre-train the \textit{Sens-BERT} with $\mathcal{X}_t^{train}$, fine-tune the calibration model ($f_\theta$) and other reference models such as regression and RF models with the paired train data, and test with the testing data. We use Adam optimizer to store and update the parameters while training and testing \cite{kingma_adam}. 

\subsection{Evaluations} \label{subsec:results}
In order to check the adaptability of \textit{Sens-BERT} to calibrate the LCS, we have performed the following three evaluations, which show that it outperforms the existing models and solves the transferability and re-calibration of calibration models for low-cost sensors under reference measurements scarcity.
\begin{enumerate}
    \item \textit{Evaluation of \textit{Sens-BERT} in the calibration of LCS} - to check whether \textit{Sens-BERT} calibrates the LCS or not
    \item \textit {Evaluation of \textit{Sens-BERT} under availability of limited paired data} -- to verify the effectiveness of \textit{Sens-BERT} under reference measurement scarcity
    \item \textit{Evaluation of \textit{Sens-BERT}transferability to other sensors} -- to find the pre-trained \textit{Sens-BERT} on some sensor can transfer to other sensors
\end{enumerate}

\begin{table}[!hbt]
\caption{Evaluation of \textit{Sens-BERT} approach for different sensors in the CAIRSENSE data set. * indicates the model that outperforms. The relative improvement with \textit{Sens-BERT} compared to the best model (RF), is shown is \%.}
\label{tab:evaluation_sens-bert}. 
        \begin{tabular}{{ccll}}
            \toprule
            \textbf{Sensor} & \textbf{Model} & $\mathbf{R^2 \uparrow}	$ & $\mathbf{RMSE} \downarrow$ \\
            \midrule
            \multirow{3}{*}{\textbf{Shinyei}}  &MLR & $0.66$   &$0.58$ \\
             & RF & $0.8$       & $0.44$  \\
              & \textit{Sens-BERT*}  & 0.9 ($12.5 \%$)    & $0.23$ ($47.7$) \\
            \midrule
            \multirow{3}{*}{\textbf{AirAssure}}      & MLR  & 0.63    & 0.61 \\
             & RF  & 0.75   & 0.5 \\
             % & Feinberg et al. \cite{cairsense_dataset} & 0.71 & NA \\
              & \textit{Sens-BERT*}   & 0.86 (17.0 \%)   & 0.3 (40\%) \\
            \midrule
            \multirow{3}{*}{\textbf{Speck}}      & MLR    &0.32  & 0.82\\
            & RF     & 0.65    & 0.58  \\
             % & Feinberg et al. \cite{cairsense_dataset} & 0.71 & NA \\
            & \textit{Sens-BERT*}   & 0.83 (27.6 \%)    & 0.32 (44.8 \%) \\
            \midrule
             \multirow{3}{*}{\textbf{TZOA}}  &MLR      &0.4  & 0.75\\
            & RF     & 0.75    & 0.49 \\
             
              & \textit{Sens-BERT*}  & 0.86 (14.6 \%)   & 0.29 (40.8 \%)\\
            \midrule
            \multirow{3}{*}{\textbf{OPCPMF}}  &MLR       &0.26   & 0.85\\
           & RF     & 0.56    & 0.66  \\

              & \textit{Sens-BERT*} & 0.81 (44.4 \%)   & 0.31 (53.3 \%) \\
            \bottomrule
        \end{tabular}
     \hfil
\end{table}
\subsubsection{Evaluation of \textit{Sens-BERT} in the calibration of LCS} \label{subsubsec:regular_evaluation} 
Here, we first pre-train the \textit{Sens-BERT} with $\mathcal{X}_t^{train}$ by following the procedure discussed in Sec. \ref{subsubsec:pre-train}. Then we fine-tune the calibration model, $f_\theta$, an LSTM-based model with $\mathcal{X}_t^{train},\mathcal{Y}_t^{train}$ by utilizing the embeddings produced by the  the pre-trained \textit{Sens-BERT}. Since the regression and RF models do not accept a bunch of samples ($\mathcal{X} \in \mathbf{R}^{M \times K} \sim \{\mathcal{X}_t\} \, \forall t $) at a time, we have flattened the sequence of samples in $\mathcal{X}$ for all the variables into a single array and passed them to the regression and RF models. For example, if the input consists of $M$ samples with $K$ variables, the flattened array consists of $M \times K$ values and is passed as a single sample to the regression and RF models. This flattening helps train the regression and RF models on the same data used for training the \textit{Sens-BERT} and ensures fair performance comparisons.  Once the training is finished, all models are tested with test data. The results of this experiment are presented in table \ref{tab:evaluation_sens-bert}, and the outperforming models are indicated with *. 

Table \ref{tab:evaluation_sens-bert} shows that \textit{Sens-BERT} outperforms the MLR and RF models for all the sensors. There is a maximum improvement of  44.4\% in $R^2$ and 53.3 \% in $RMSE$ for the OPCPMF Sensor. At a minimum, there is a 12.5\% improvement in $R^2$ for Shinyei and a 40\% improvement in $RMSE$ for AirAssure sensors. Therefore,  $\textit{Sens-BERT}$ can be adopted for calibrating low-cost air quality sensors. 

\subsubsection{Evaluation of \textit{Sens-BERT} under availability of limited paired data}

\begin{figure}[!hbt]
    \centering
    \includegraphics[width = 0.8\textwidth]{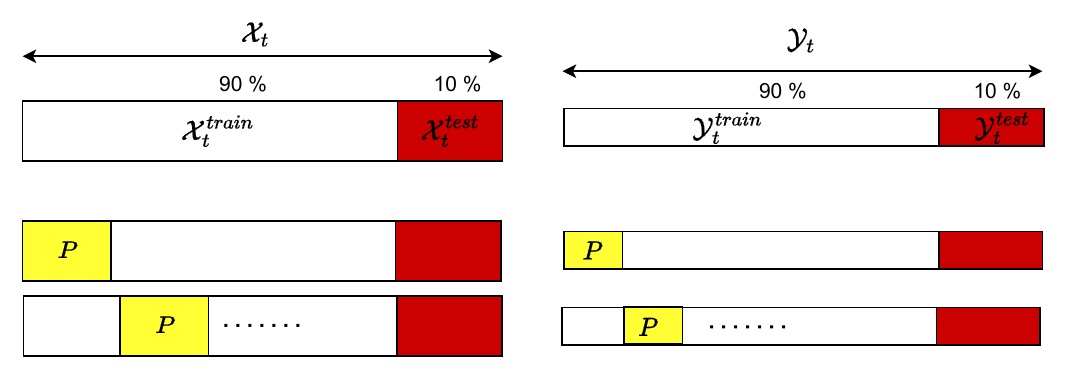}
    \caption{Splitting of training data into chunks to test the calibration models under limited pared data}
    \label{fig:datasplit}
\end{figure}

The initial assumption of this experiment is that there is a scarcity of reference station observations ($\mathcal{Y}_t$). In contrast to the experiment in Sec. \ref{subsubsec:regular_evaluation}, Let's assume only a few measurements in $\mathcal{X}_t^{train}$ has corresponding reference station observations in $\mathcal{Y}_t^{train}$. Then it is not possible to train the MLR and RF models on entire train data since they need paired data. However, \textit{Sens-BERT} has the advantage of learning the characteristics of $\mathcal{X}_t$ without having $\mathcal{Y}_t$ observations. To implement the limited paired data experiments, we assume that only  $P$ \% of samples in $\mathcal{X}_t^{train}$ have corresponding reference station observations. We then train the calibration models with that $P$ \% data and test with the testing data. We repeat this by considering other $P$ \% samples in $\mathcal{X}_t^{train}$ until it covers complete data of $\mathcal{X}_t^{train}$ as shown in Fig. \ref{fig:datasplit}. Finally, we average the $R^2$ and $RMSE$ values obtained in testing for all the sets, and we repeat this experiment by increasing the $P$ value, such that it equals 100\%, represents the experiment in Sec \ref{subsubsec:regular_evaluation}. This procedure helps test models' calibration efficiency with limited paired data since we pass only limited paired data. Yet, we make models to expose a complete range of values.

\begin{figure}[!hbt]
    \begin{minipage}[t]{1\textwidth}
    \includegraphics[scale = 0.47]{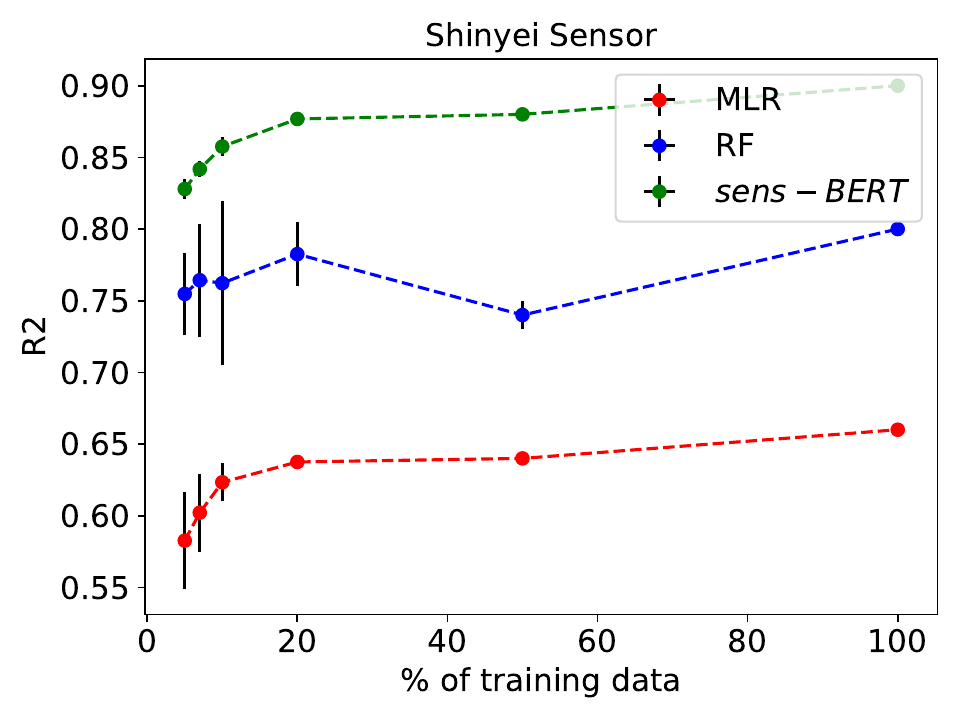}
    % \subcaption{  }
    % \end{minipage}%
    \hfill
% \begin{minipage}[t]{0.45\textwidth}
    \includegraphics[scale = 0.47]{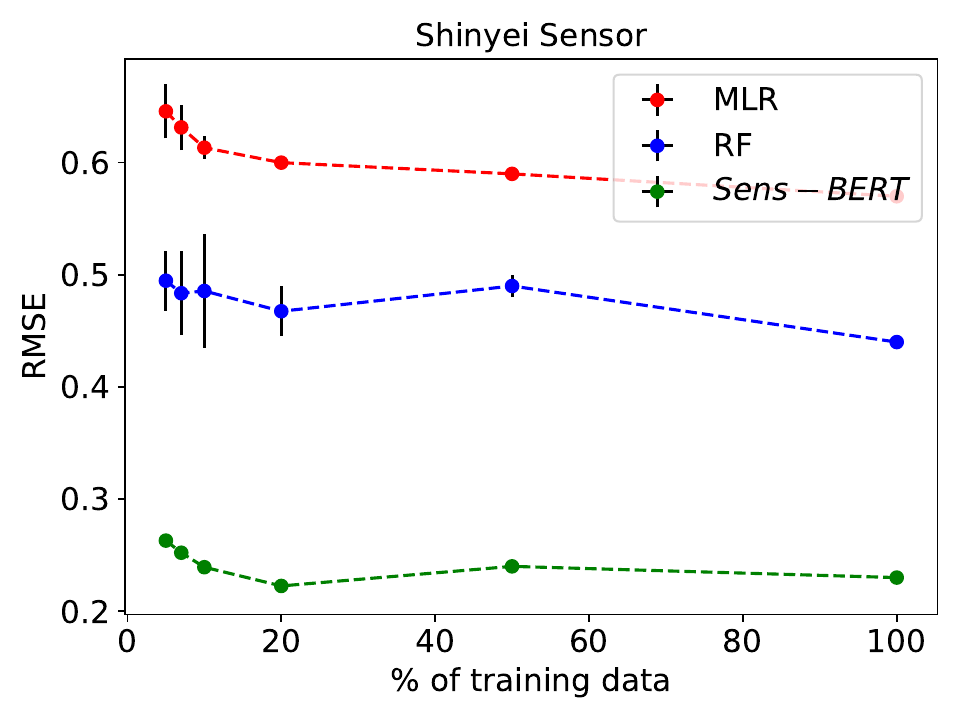}
    \subcaption{ $R^2$ and $RMSE$ for Shinyei sensor}
    \end{minipage}%
    \\
     \begin{minipage}[t]{1\textwidth}
    \includegraphics[scale = 0.47]{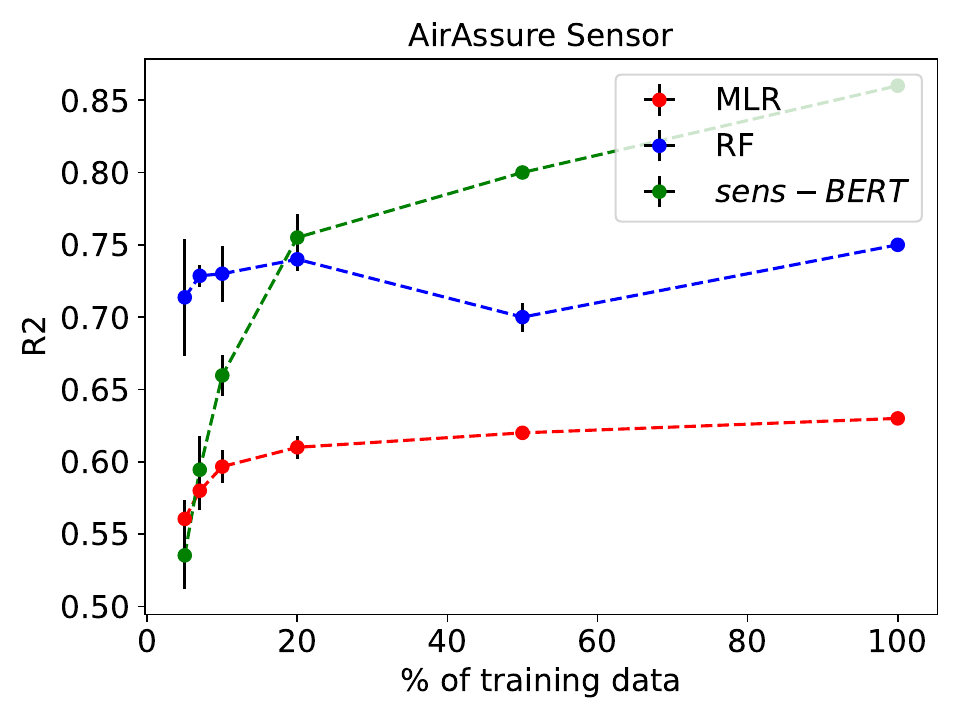}
    % \subcaption{}
    % \end{minipage}%
    \hfill
% \begin{minipage}[t]{0.45\textwidth}
    \includegraphics[scale = 0.47]{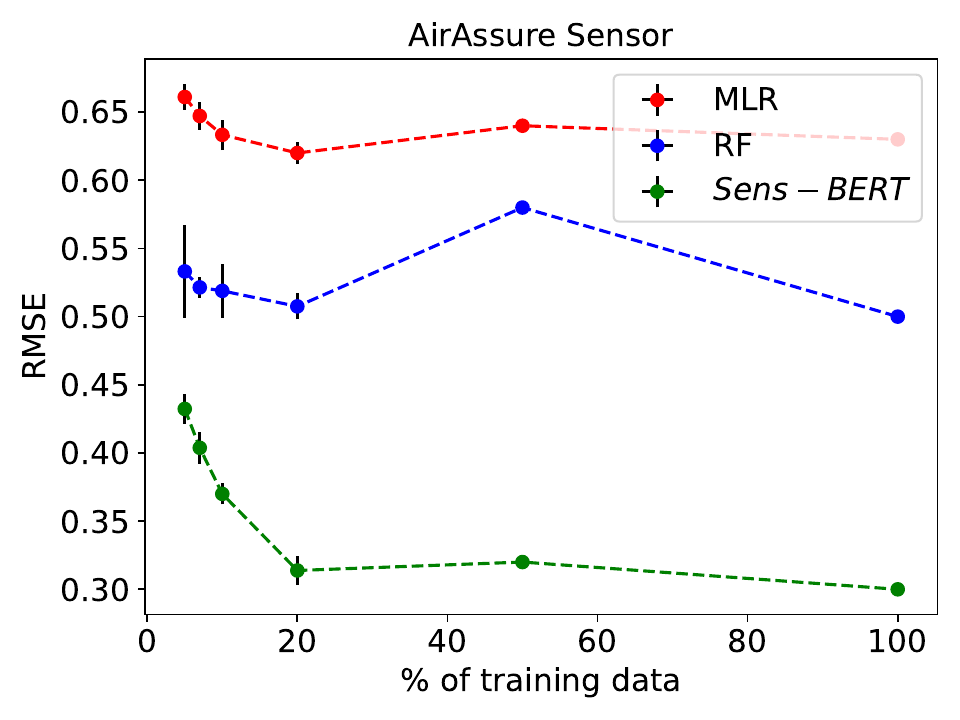}
    \subcaption{ $R^2$ and $RMSE$  for AirAssure sensor}
    \end{minipage}%
    \caption{Mean $R^2$ and $RMSE$ values of different calibration models with one standard deviation with varying amounts of training data }
    \label{fig:limiteddata_exp}
\end{figure}

We expect the calibration accuracy of MLR and RF models increases as the $P$ increases since these models can learn better with more paired data. In the case of calibration with \textit{Sens-BERT}, we expect that it maintains the calibration accuracy even the $P$ values decrease since it learns the characteristics of $\mathcal{X}_t$ in pre-training and does not require more paired data to calibrate LCS. However, a too-low $P$ value may also influence the calibration performance of \textit{Sens-BERT} since a neural network architecture needs minimum data. We tested Shinyei and AirAssure sensors in this experiment, and the mean $R^2$ and $RMSE$ values with one standard deviation are presented in Fig. \ref{fig:limiteddata_exp}. Fig. \ref{fig:limiteddata_exp} shows that For Shinyei Sensors, the \textit{Sens-BERT} based calibration outperforms the MLR and RF models. However, in the case of AirAssure sensors, \textit{Sens-BERT} require at least 20 \% of the train data to fine-tune the LSTM-based calibration model ($f_\theta$). After that, it outperforms both the MLR and RF models. Due to computational constraints, we have experimented on two sensors, and we expect the \textit{Sens-BERT} based calibration can work for other sensors under reference measurement scarcity.

\subsubsection{Evaluation of transferability of \textit{Sens-BERT} to other sensors}
In this experiment, we first adopt the pre-trained models mentioned in Sec. \ref{subsubsec:regular_evaluation}. Then we fine-tune the pre-trained models with other sensors' train data and test with the test data.  For example, if a model is pre-trained on the data of sensor $K1$, then it is fine-tuned with the train data of sensor $K2$ and tested with the testing data of the $K2$. This process helps check the transferability of the models trained on one sensor to another.  The experimental results for training on the AirAssure sensor and testing on Shinyei sensors are presented in Table \ref{tab:transferability}.

% \begin{table*}[!hbt]
% \caption{Transferability check of calibration models. }
% \label{tab:transferability}
%     \begin{subtable}[t]{0.5\textwidth}
%     \caption{ Trained on AirAssure}
%         \begin{tabular}{{|c|c|c|c|c|c|}}
%             \hline
%            {Model} & {tested on}  & $R^2$ & \% in change in $R^2$  & $RMSE$ & \% change in $ RMSE$     \\      
%             \hline 
%             MLR & \multirow{3}{*}{Speck} & -0.04 & & 1.02&  \\
%             RF & & 0.02 & & 0.99 &\\
%             \textit{Sens-BERT} &  & 0.7 & &0.36&   \\
%             \hline
%             MLR & \multirow{3}{*}{Shinyei} & 0.48 & & 0.72&  \\
%             RF & & 0.65 & & 0.59 &\\
%             \textit{Sens-BERT} &  & 0.8 & &0.34&   \\
%             \hline
            
%             MLR & \multirow{3}{*}{TZOA} & 0.13 & & 0.9&  \\
%             RF & & 0.53 & & 0.66 &\\
%             \textit{Sens-BERT} &  & 0.73 & &0.32&   \\
%             \hline
%         \end{tabular}
%     \end{subtable}
%      \hfil
%      \begin{subtable}[t]{0.5\textwidth}
%     \caption{ Trained on Speck}
%         \begin{tabular}{{|c|c|c|c|c|c|}}
%             \hline
%            {Model} & {tested on}  & $R^2$ & \% in change in $R^2$  & $RMSE$ & \% change in $ RMSE$     \\      
%             \hline 
%             MLR & \multirow{3}{*}{Shinyei} & 0.48 & & 0.72&  \\
%             RF & & 0.65 & & 0.59 &\\
%             \textit{Sens-BERT} &  & 0.8 & &0.34&   \\
%             \hline
%         \end{tabular}
%     \end{subtable}
%      \hfil
% \end{table*}

\begin{table} 
\caption{ Testing transferability of calibration models }
\label{tab:transferability}
\begin{minipage}[b]{0.4\textwidth}
\subcaption{ Trained on AirAssure}
% \resizebox{\textwidth}{!}{
\begin{tabular}{clll}
\toprule
\textbf{Model}                      & \textbf{tested on} & $R^2$ $\uparrow$ & $RMSE$ $\downarrow$ \\
\midrule
\multirow{4}{*}{\textbf{MLR}}       & Speck         & 0.04   &  1.02    \\
                           & TZOA    & 0.13   & 0.9     \\
                           & Shinyei &0.48  & 0.72\\
                           &  AirAssure& 0.66 & 0.58 \\
\midrule
\multirow{4}{*}{\textbf{RF}}        & Speck         & 0.02   &  0.99    \\
                           & TZOA    &0.53    & 0.66    \\
                           & Shinyei &0.65  &0.59 \\
                           &  AirAssure&0.75  & 0.5 \\
\midrule
\multirow{4}{*}{\textbf{\textit{Sens-BERT}}} & Speck         & 0.7   &  0.36    \\
                           & TZOA    &0.73    &0.32      \\
                           & Shinyei & 0.8 & 0.34\\
                           &  AirAssure& 0.86 & 0.3 \\
\bottomrule
\end{tabular}
% }
\end{minipage}%
\quad\quad\quad
\begin{minipage}[b]{0.4\textwidth}
\subcaption{ Trained on Speck}
% \resizebox{\textwidth}{!}{
\begin{tabular}{clll}
\toprule
\textbf{Model}                      & \textbf{tested on} & $R^2$ $\uparrow$ & $RMSE$ $\downarrow$ \\
\midrule 
\multirow{4}{*}{\textbf{MLR}}       & Speck         & 0.32   &  0.82    \\
                           & TZOA    & -2.29   & 1.75     \\
                           & Shinyei &-0.19   & 1.09\\
                           & AirAssure& 0.27 & 0.85 \\
\midrule
\multirow{4}{*}{\textbf{RF}}        & Speck         & 0.65  &  0.58    \\
                           & TZOA    &0.01    & 0.96    \\
                           & Shinyei &0.52  &0.69 \\
                           & AirAssure&0.19 & 0.9 \\
\midrule
\multirow{4}{*}{\textbf{\textit{Sens-BERT}}} & Speck         & 0.83   &  0.29    \\
                           & TZOA    &0.73    &0.35      \\
                           & Shinyei & 0.78 & 0.34\\
                           & AirAssure& 0.7 & 0.39 \\
\bottomrule
\end{tabular}
% }
\end{minipage}%
\end{table}

% \section{Ablations}
% \subsection{Performance of \textit{Sens-BERT} with different sequence lengths}
% \subsection{Testing the calibration models with different combinations of input variables }
\section{Conclusions and future scope }
This work proposes \textit{Sens-BERT}, a BERT-based deep learning calibration approach for calibrating low-cost sensors for air quality monitoring. \textit{Sens-BERT} outperforms the regression and random forest machine learning algorithms, which are extensively used in low-cost sensor calibration. We empirically validated the calibration performance of  \textit{Sens-BERT} on different low-cost sensors from the CAIRSENSE data set, available on US Environmental Protection Agency (USEPA) data website. \textit{Sens-BERT} enables the re-calibration, the frequent calibration after sensor deployment to overcome the calibration drift, with limited paired data. We tested the performance of \textit{Sens-BERT} under limited paired data by varying   Further, {Sens-BERT} trained on one sensor can be transferred to other sensors. However, we tested the transferability on sensors working on the same sensing principle, that means if \textit{Sens-BERT} is trained on a sensor works on an optical principle, we transferred the trained \textit{Sens-BERT} to other sensors that also work on optical principle. Checking transferability between sensors works on different sensing principles is our future work. 

In general performance of the BERT-based model improves if it learns a good representation of data. On that note, we expect that \textit{Sens-BERT} trained on a substantial amount of low-cost sensor measurements curated from various data sets and websites makes it more powerful. However, it needs more computational capacity, which is a constraint for us now.  

%  \textit{Sens-BERT} address calibration accuracy, re-calibration, and calibration tran sfer, the three main issues of low-cost sensor calibration. In the caseo o In contrast to the existing calibration works \textit{Sens-
% BERT} first learns the characteristics of LCS measurements without using reference station observations. This 

%   the well-known machine learning algorithms such as regression and random forest, which are extensively used for calibrating LCS. Since the \textit{Sens-BERT} It addresses the calibration model transferability 

\bibliographystyle{unsrt}
\bibliography{reference}

\end{document}